# Effect of electron-phonon coupling on thermal transport in metals: a Monte Carlo approach for solving the coupled electron-phonon Boltzmann transport equations

Jie Peng [1, a)], W. Ryan Deskins [1], Maria Helena Braga[2], and Anter El-Azab [1,a)]

[1] Materials Engineering, Purdue University, West Lafayette, Indiana 47907, USA
[2] LAETA, Engineering Physics Department, Engineering Faculty, University of Porto, R. Dr Roberto Frias s/n, 4200-465 Porto, Portugal
a) Author to whom correspondence should be addressed: jpeng02@villanova.edu

**ABSTRACT**

In this work, the effect of electron-phonon (e-ph) coupling on both electron and phonon transport of metals is investigated via first principles calculations. A Monte-Carlo (MC) approach for solving the coupled electron-phonon Boltzmann transport equations is developed to investigate thermal conductivity of metals. In this approach, the anisotropic electron band structure, phonon dispersion in the full Brillouin zone, and mode-dependent thermal relaxation time of electrons and phonons are calculated from first principles. Using this approach, MC simulations of coupled e-ph thermal transport at different temperatures in $\alpha$-U and Ag are performed. These two materials were selected as a way to demonstrate the applicability of the method in different fields. The results indicate that the electron relaxation time due to phonon scattering is orders of magnitude smaller than the phonon relaxation time due to electron scattering. The results also show that, in phonon thermal transport, the impact of ph-e scattering is almost negligible and the ph-ph scattering dominates phonon transport. At high temperature, the electrons dominate thermal transport in both $\alpha$-U and Ag. However, at low temperature, the phonon contribution to the total thermal conductivity of $\alpha$-U is significant. Moreover, the Lorenz ratio deviates from the Sommerfeld value at low to intermediate temperatures, where the Wiedemann-Franz law is not applicable. Finally, we show that the Ag electronic thermal conductivity shows a stronger size effect than its phonon thermal conductivity.

# 1 Introduction

There have been numerous studies on thermal transport in metals in the past few decades [1, 2, 3, 4]. Typically, electrons are treated as dominating thermal transport, while phonons are assumed to make negligible contribution. Therefore, many studies only consider electronic thermal conductivity in metals. However, there has been a growing interest in separating and comparing the electron and phonon contributions to thermal conductivity in metals. Recent studies have found that phonon heat conduction in metals plays an important role in a variety of energy transfer processes including interfacial thermal transport, laser heating, and heat-assisted magnetic recording [5, 6, 7]. Experimentally, direct measurement of the electronic and phonon thermal conductivity is difficult. In general, the electrical conductivity is measured and then used to determine the electronic thermal conductivity through the Wiedemann-Franz law [4]. However, the Wiedemann-Franz law has been shown to fail at low and intermediate temperatures due to inelastic electron scattering events [4, 8]. From a theoretical modeling perspective, how to quantify electronic and phonon thermal conductivity of metals has become an important problem.

The e-ph coupling plays an important role in both electron and phonon thermal transport. While the effect of this coupling on electrical transport properties has been extensively studied, its effect on thermal transport has only gained attention recently. In the early works by Leibfried and Schlomann [9], Klemens [10], and Slack [11], the e-ph scattering is neglected from the models of phonon thermal transport. Later works that incorporate e-ph coupling in the conductivity models mostly assume free electrons and only consider their interactions with long-wavelength acoustic phonons [2, 12, 3]. Nevertheless, such assumption fails to account for the e-ph coupling of individual phonon modes and electron states in the full Brillouin zone (BZ). In recent decades, the development of first-principles density functional theory (DFT) approaches to evaluate phonon-phonon (ph-ph) [13, 14] and e-ph interactions [6, 15, 16] has enabled calculation of e-ph scattering strength in the full BZ. The DFT method provides an accurate approach to model thermal conductivity of metals.

Both electron and phonon thermal transport are governed by the Boltzmann transport equation (BTE) [4, 17, 16]. Methods for solving the BTE include the non-transport solution, where the BTE itself is converted into a formula for conductivity, [18, 19, 20], the lattice Boltzmann method [21, 22], and the discrete ordinate method [23, 24]. Among these methods, the stochastic MC method provides an approach to solve thermal problems in 3D and for devices with complex geometry [25, 26, 27]. Furthermore, it can treat the details of scattering events at the microscopic scale and be applied to study anisotropic energy transport. However, previous MC studies for solving the coupled e-ph BTEs are based on overly simplified models such as isotropic phonon dispersion, parabolic electron band structure, and empirical e-ph relaxation time [28, 29, 30]. For these reasons, a MC framework that incorporates the full Brillouin zone (BZ) electron and phonon scattering properties to study thermal transport problems is desirable.

It is to be noted here that solving the coupled e-ph BTEs for both thermal and electrical transport entails the same details. This exercise is promising from the perspective of nanoscale device applications. The e-ph interaction determines the intrinsic behavior of the electrical conductivity in semiconductors, limiting the assisted mobility in 2D semiconductors (e.g., $MoS_2$, $WSe_2$, $SnS_2$…) while being a significant source of decoherence, causing scattering and joule dissipation in electronic devices such as field effect transistors (FET) [31, 32, 33, 34]. Nevertheless, the e-ph coupling may strategically convert into a positive feature by reaching strong coupling regimes in phonon cavities in charge qubits [35], phononic or acoustic metamaterials [36], and metallic nanocrystals [37]. Conversely, it is recognized that an 'ideal' thermoelectric (TE) allows electrons to flow freely, like through the surface of a metal or, better, a superconductor. As glass, an ideal TE scatters phonons heavily. The idea of designing such a material – an 'electron-crystal and phonon-glass' [38, 39] – for a high thermoelectric figure of merit has boosted studies in the field of TEs. One such study is related to relaxor ferroelectrics, where localized phonon modes provide phonon-glass properties due to the enhanced levels of phonon scattering owing to nanopolar regions [40, 41].

Although the best TEs were also some of the first materials to be recognized as topologic insulators (TIs), the origin of the e-ph coupling in these materials remains hard to investigate, as it is hard to isolate the contribution of the subsurface phonons of the insulator bulk experimentally. However, it is worth exploring many-body interactions in TIs as they are likely the future of dissipationless electronics, energy harvesting and storage devices, and quantum computing [42, 43, 44, 45, 46]. To understand and, subsequently, tailor responses in artificial materials and devices, it is vital to accurately predict e-ph coupling behavior in metals, semiconductors, and insulators, especially at the nanoscale.

In this work, we introduce a MC approach to solve the coupled e-ph BTE and investigate the effect of e-ph scattering on thermal transport in $\alpha$-U and Ag. The phonon dispersion, electron band structure, and relaxation times of electrons and phonons are all calculated from first principles. A 'deviational' MC scheme for solving the coupled e-ph BTE is developed. The rest of this paper is organized as the following. In Sec. 2, the theoretical formulation of coupled e-ph BTE and the MC simulation steps are described. The DFT calculated electron and phonon properties are presented in Sec. 3. Discussion on the thermal transport properties of $\alpha$-U and Ag as well as the effect of e-ph coupling obtained from MC simulations are detailed in Sec. 4.

## 2 Methodology

### 2.1 Coupled e-ph Boltzmann transport equation

In the absence of an external field, the coupled e-ph BTEs are expressed as [4]

$$\begin{cases} \dfrac{\partial f_{\alpha k}}{\partial t} = -v_{\alpha k}^{\text{e}} \nabla_r f_{\alpha k} + \left(\dfrac{\partial f_{\alpha k}}{\partial t}\right)_s \\ \dfrac{\partial n_{\beta q}}{\partial t} = -v_{\beta q}^{\text{ph}} \nabla_r n_{\beta q} + \left(\dfrac{\partial n_{\beta q}}{\partial t}\right)_s \end{cases}, \quad (1)$$

where $f_{\alpha k} = f_{\alpha k}(r,t)$ is the electron distribution function representing the occupation number at wavevector index $k$, electron band index $\alpha$, spatial position $r$, time $t$, and $n_{\beta q} = n_{\beta q}(r,t)$ the phonon distribution function representing the occupation number at wavevector index $q$ and phonon branch index $\beta$, $v_{\alpha k}^{\text{e}}$ is the electron velocity, and $v_{\beta q}^{\text{ph}}$ is the phonon group velocity. The scattering terms $(\partial f_{\alpha k}/\partial t)_s$ and $(\partial n_{\beta q}/\partial t)_s$ represent the alteration of distribution function due to electron and phonon scattering processes.

To evaluate the scattering terms, the widely used relaxation time approximation (RTA) is used to decouple the e-ph BTE. Full decoupling requires evaluation of nonequilibrium electron and phonon states, which is beyond the scope of this work. The RTA assumes that the electron and phonon distribution functions return to their local thermal equilibrium at a rate given by [4, 18]

$$\begin{cases} \left(\dfrac{\partial f_{\alpha k}}{\partial t}\right)_s = \dfrac{f_{\alpha k} - f_{\alpha k}^0}{\tau_{\alpha k}^{\text{e}}} \\ \left(\dfrac{\partial n_{\beta q}}{\partial t}\right)_s = \dfrac{n_{\beta q} - n_{\beta q}^0}{\tau_{\beta q}^{\text{ph}}} \end{cases}, \quad (2)$$

where $\tau_{\alpha k}^{e}$ and $\tau_{\beta q}^{ph}$ are the electron and phonon relaxation times, respectively, and $f_{\alpha k}^{0}$ and $n_{\beta q}^{0}$ are the equilibrium distribution functions of electrons and phonons, given by the Fermi-Dirac and Bose-Einstein distributions, respectively [47]

$$\begin{cases} f_{\alpha k}^{0} = \dfrac{1}{\exp\left(\dfrac{\epsilon_{\alpha k} - \mu}{k_B T_e}\right) + 1} \\ n_{\beta q}^{0} = \dfrac{1}{\exp\left(\dfrac{\hbar \omega_{\beta q}}{k_B T_{ph}}\right) - 1} \end{cases}, \qquad (3)$$

where $\epsilon_{\alpha k}$ is the electron band energy, $\omega_{\beta q}$ is the phonon frequency, $\mu$ is the chemical potential for electrons, $T_e$ and $T_{ph}$ are the electron and phonon temperature, respectively, $\hbar$ is the reduced Planck constant, and $k_B$ is the Boltzmann constant. The chemical potential is very close to the Fermi energy $\epsilon_F$ at the temperatures considered in this work, so that $\mu = \epsilon_F$ is assumed [47].

Only scattering events among electrons and phonons are considered in this work; that is defect scattering is not considered. For the electronic thermal conductivity of metals, it is found that the contribution from the scattering of electrons by electrons (e-e scattering) is much smaller compared to the contribution from scattering of electrons by phonons (e-ph scattering) [8, 48]. Therefore, the electron relaxation time only includes e-ph scattering while both ph-ph and phonon scattering by electrons (ph-e scattering) are included in the phonon relaxation time calculations. The overall electron and phonon relaxation times are calculated via Matthiessen's rule [8, 49]

$$\begin{cases} \dfrac{1}{\tau_{\alpha k}^{e}} = \dfrac{1}{\tau_{\alpha k}^{e-ph}} \\ \dfrac{1}{\tau_{\beta q}^{ph}} = \dfrac{1}{\tau_{\beta q}^{ph-ph}} + \dfrac{1}{\tau_{\beta q}^{ph-e}} \end{cases}, \qquad (4)$$

where $\tau_{\alpha k}^{e-ph}$ is the e-ph relaxation time of electron and $\tau_{\beta q}^{ph-ph}$ and $\tau_{\beta q}^{ph-e}$ are the ph-ph and ph-e relaxation times of phonons, respectively.

By combining Eq. (1) and (2) we can rewrite the coupled e-ph BTEs as

$$\begin{cases} \dfrac{\partial f_{\alpha k}}{\partial t} = -v^{e}_{\alpha k}\nabla_r f_{\alpha k} + \dfrac{f_{\alpha k} - f^{0}_{\alpha k}}{\tau^{e}_{\alpha k}} \\ \dfrac{\partial n_{\beta q}}{\partial t} = -v^{ph}_{\beta q}\nabla_r n_{\beta q} + \dfrac{n_{\beta q} - n^{0}_{\beta q}}{\tau^{ph}_{\beta q}} \end{cases} \quad (5)$$

## 2.2 Deviational Monte Carlo formulation

The implemented MC scheme seeks the solution of Eqs. (5). A recently developed deviational form which can reduce the statistical uncertainty and the computational cost of the traditional MC form is adopted [50, 51]. Below we will present the definition of macroscopic variables that are reformulated into the deviational form.

When considering thermal transport by electrons, it is not feasible to track the electron distribution function in the entire wavevector space due to the large number of electrons. In fact, only electrons around the Fermi level within a width of a few thermal energy units ($\sim k_B T_e$) can respond to thermal perturbation [4]. Therefore, only a small population of the entire electronic band spectrum needs to be considered in the MC scheme. We start from an electron system for which the electron distribution function $f$ is known while the local temperature is not yet defined. Its excitation energy density is given by the difference between its energy density and the ground state energy density as

$$\begin{aligned} \Delta E_e(f) &= E_e(f) - E_e(T=0) \\ &= \int_{-\infty}^{\infty} f \epsilon D_e(\epsilon) d\epsilon - \int_{-\infty}^{\epsilon_F} \epsilon D_e(\epsilon) d\epsilon \\ &= \int_{-\infty}^{\epsilon_F} (\epsilon_F - \epsilon)(1-f) D_e(\epsilon) d\epsilon + \int_{\epsilon_F}^{\infty} (\epsilon - \epsilon_F) f D_e(\epsilon) d\epsilon, \end{aligned} \quad (6)$$

where $D_e(\epsilon)$ is the electron density of states per unit volume. Here we introduce the deviational electron energy density $E_e^d$ defined at a reference temperature $T_{\text{ref}}$ [51]

$$E_e^d = \Delta E_e(f) - \Delta E_e(T_{\text{ref}}) = \int_{-\infty}^{\infty} |\epsilon - \epsilon_F| g_e^d D_e(\epsilon) d\epsilon, \tag{7}$$

where $g_e^d = \begin{cases} -f(\epsilon) + f^0(\epsilon, T_{\text{ref}}), \epsilon \leq \epsilon_F \\ f(\epsilon) - f^0(\epsilon, T_{\text{ref}}), \epsilon \geq \epsilon_F \end{cases}$ is the deviational electron distribution function. Using Eq. (7), the local temperature can be determined by assuming that the deviational energy density is equal to that at an equivalent equilibrium temperature $T_{\text{loc}}$ such that we may write

$$E_e^d = \Delta E_e(f(T_{\text{loc}})) - \Delta E_e(T_{\text{ref}}) = \int_{-\infty}^{\infty} |\epsilon - \epsilon_F| g_{\text{loc}}^d D_e(\epsilon) d\epsilon, \tag{8}$$

where $g_{\text{loc}}^d = f_0(\epsilon, T_{\text{loc}}) - f_0(\epsilon, T_{\text{ref}})$ is the deviational electron equilibrium distribution. Now, the electron BTE in Eq. (5) is rewritten into a deviational form as

$$\frac{\partial g_e^d}{\partial t} = -v^e \nabla_r g_e^d - \frac{g_e^d - g_{\text{loc}}^d}{\tau^e}, \tag{9}$$

where the subscripts are dropped since an integration over the electron band is performed, as shown in Eq. (7). Similarly, for a phonon system with a distribution function $n$, its deviational energy density is defined as

$$E_{\text{ph}}^d = E_{\text{ph}}(n) - E_{\text{ph}}(T_{\text{ref}}) = \int_0^{\omega_{max}} \hbar\omega D_{\text{ph}}(\omega) |n - n^0(\omega, T_{\text{ref}})| d\omega, \tag{10}$$

where $D_{\text{ph}}(\omega)$ is the phonon density of states per unit volume, and $\omega_{max}$ is the highest phonon frequency. The local temperature can be determined by assuming that the deviational energy density is equal to that at an equivalent equilibrium temperature $T_{\text{loc}}$,

$$E_{\text{ph}}^d = \int_0^{\omega_{max}} \hbar\omega D_{\text{ph}}(\omega) |n^0(\omega, T_{\text{loc}}) - n^0(\omega, T_{\text{ref}})| d\omega. \tag{11}$$

Let $h_{\text{ph}}^d = n - n^0(\omega, T_{\text{ref}})$ be defined as the phonon deviational distribution function. The phonon BTE in Eq. (5) is then rewritten into its deviational form as

$$\frac{\partial h_{\mathrm{ph}}^d}{\partial t} = -v^{\mathrm{ph}}\nabla_r h_{\mathrm{ph}}^d - \frac{h_{\mathrm{ph}}^d - h_{\mathrm{loc}}^d}{\tau^{\mathrm{ph}}}, \tag{12}$$

where $h_{\mathrm{loc}}^d = n^0(\omega, T_{\mathrm{loc}}) - n^0(\omega, T_{\mathrm{ref}})$ is the deviational phonon equilibrium distribution. By combining Eq. (9) and Eq. (12), the coupled e-ph BTE is decoupled into a deviational form as

$$\begin{cases} \dfrac{\partial g_{\mathrm{e}}^d}{\partial t} = -v^{\mathrm{e}}\nabla_r g_{\mathrm{e}}^d - \dfrac{g_{\mathrm{e}}^d - g_{\mathrm{loc}}^d}{\tau^{\mathrm{e}}} \\ \dfrac{\partial h_{\mathrm{ph}}^d}{\partial t} = -v^{\mathrm{ph}}\nabla_r h_{\mathrm{ph}}^d - \dfrac{h_{\mathrm{ph}}^d - h_{\mathrm{loc}}^d}{\tau^{\mathrm{ph}}} \end{cases}, \tag{13}$$

By solving for the distribution function from Eq. (13), the phonon and electron heat flux can be determined as

$$\begin{cases} \vec{q}_{\mathrm{e}} = \displaystyle\int_{-\infty}^{\infty} \vec{v}^{\mathrm{e}} |\epsilon - \epsilon_F| g_{\mathrm{e}}^d D_{\mathrm{e}}(\epsilon) d\epsilon \\ \vec{q}_{ph} = \displaystyle\int_{0}^{\omega_{max}} \vec{v}^{\mathrm{ph}} \hbar\omega h_{\mathrm{ph}}^d D_{\mathrm{ph}}(\omega) d\omega \end{cases}, \tag{14}$$

## 2.3 Monte Carlo scheme

In this section, we will introduce the MC scheme for solving the deviational, decoupled e-ph BTE. In our MC scheme, all electrons and phonons are sampled from a grid in wavevector space. Let the total number of k-points in the electron wavevector space and q-points in the phonon wavevector space be $N_k$ and $N_q$, respectively. There are a total number of $N_k N_\alpha$ electron states and $N_q N_\beta$ phonon modes, with $N_\alpha$ being the number of electron bands and $N_\beta$ the number of phonon branches. The sampled electrons and phonons are allowed to diffuse and scatter in the simulation domain while statistics are collected and used to extract thermal properties. In Figure 1(a), we show a simple flow chart of the MC scheme.

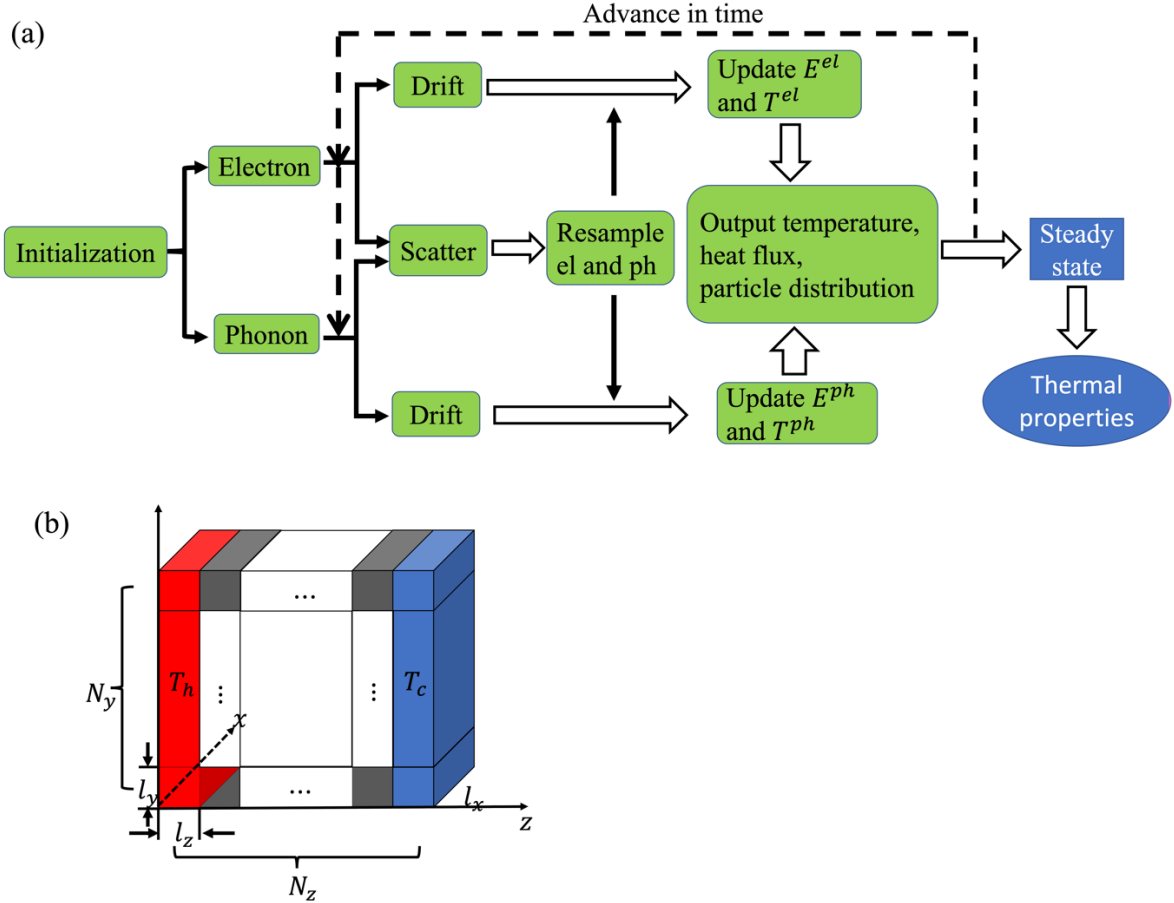

Figure 1. (a) A flowchart of the MC scheme for solving the coupled e-ph BTE. The MC simulation starts by initializing electrons and phonons according to the band structure and dispersion data obtained from DFT, respectively. The simulation steps in time with each timestep being composed of a drift and scattering phase. After each drift process, the energy of each cell is determined and used to update its pseudo-temperature. The temperature, particle distribution, and heat flux data are then recorded. After reaching thermal equilibrium or steady state, depending on the boundary conditions, the thermal properties are analyzed. Details of each step in the MC scheme are presented in this section. (b) A sketch of the three-dimensional (3D) simulation domain used in this work. The rectangular domain is discretized into $N_x \times N_y \times N_z$ cells, each having dimension $l_x \times l_y \times l_z$. This configuration can be generalized into any 3D geometry with the desired boundary conditions.

a) Simulation setup

Before the simulation begins, we must first choose the shape and boundary conditions of the simulation domain. In this work, we consider the case where a temperature gradient is imposed along one direction. As shown in Figure 1(b), a simple rectangular shaped domain is considered. The cell temperatures at the two ends of the simulation domain in the z-direction are fixed at a hot temperature $T_\text{h}$ and a cold temperature $T_\text{c}$, respectively. The temperature difference, therefore, produces a thermal gradient along the z-direction that drives the electron and phonon transport. The boundary conditions in the other two directions can vary depending on the desired simulation, e.g., periodic, reflective, etc.

b) Initialization

The algorithm starts by choosing a reference temperature $T_{\text{ref}}$ from which the deviational distributions will be sampled. The smaller the deviation from the equilibrium state, the smaller the number of deviational particles required for a given statistical uncertainty, and the more efficient the simulation will be. Thus, a natural choice of $T_{\text{ref}}$ is the expected equilibrium temperature. If the system is initialized at a temperature $T_0$, the initial number of deviational phonons per unit volume is determined as

$$N_{\text{ph}}^{\text{ini}} = \sum_{q=1}^{N_q} \sum_{\beta=1}^{N_\beta} \frac{1}{N_q V_u} |n^0(\omega_{\beta q}, T_0) - n^0(\omega_{\beta q}, T_{\text{ref}})|, \qquad (15)$$

where $V_u$ is the unit cell volume. For electrons, a Fermi window $[\epsilon_1, \epsilon_2]$ is chosen to only simulate electrons that can be thermally perturbed. A typical choice of the Fermi window $\epsilon_1 = \epsilon_F - 10k_BT$ and $\epsilon_2 = \epsilon_F + 10k_BT$ is used [50, 51]. Let us denote the number of bands in the Fermi window by $N'_\alpha$. The initial number of deviational electrons per unit volume is determined as

$$N_{\text{e}}^{\text{ini}} = \sum_{k=1}^{N_k} \sum_{\alpha=1}^{N'_\alpha} \frac{1}{N_k V_u} \left| \frac{1}{\exp\left(\frac{|\epsilon_{\alpha k} - \epsilon_F|}{k_B T_0}\right) + 1} - \frac{1}{\exp\left(\frac{|\epsilon_{\alpha k} - \epsilon_F|}{k_B T_{\text{ref}}}\right) + 1} \right|. \qquad (16)$$

For a rectangular cell with dimension $l_x \times l_y \times l_z$ in the simulation domain, the number of electrons and phonons are given by

$$\begin{cases} N_e^0 = \dfrac{1}{W} \sum_{k=1}^{N_k} \sum_{\alpha=1}^{N'_\alpha} \dfrac{l_x \times l_y \times l_z}{N_k V_u} \left| \dfrac{1}{\exp\left(\dfrac{|\epsilon_{\alpha k} - \epsilon_F|}{k_B T_0}\right) + 1} - \dfrac{1}{\exp\left(\dfrac{|\epsilon_{\alpha k} - \epsilon_F|}{k_B T_{\text{ref}}}\right) + 1} \right|, \\ N_{\text{ph}}^0 = \dfrac{1}{W} \sum_{q=1}^{N_q} \sum_{\beta=1}^{N_\beta} \dfrac{l_x \times l_y \times l_z}{N_q V_u} |n^0(\omega_{\beta q}, T_0) - n^0(\omega_{\beta q}, T_{\text{ref}})| \end{cases} \quad (17)$$

where $W$ is a weighting factor to reduce the number of particles. Thus, each particle in the MC simulation represents an ensemble of $W$ electrons/phonons with same energy and wavevector.

After determining the number of electrons and phonons in each cell, the next step is to initialize electron and phonon properties based on the electronic band structure and phonon dispersion. The discretization of the BZ provides a finite set of wavevectors, and so, electrons and phonons are sampled accordingly. Using the phonon dispersion, a cumulative probability distribution function (CPDF) of phonon wavevectors is calculated as

$$F_i^{\text{ph}} = \frac{\sum_{q=1}^{i} \sum_{\beta=1}^{N_\beta} [n^0(\omega_{\beta q}, T_0) - n^0(\omega_{\beta q}, T_{\text{ref}})]}{\sum_{q=1}^{N_q} \sum_{\beta=1}^{N_\beta} [n^0(\omega_{\beta q}, T_0) - n^0(\omega_{\beta q}, T_{\text{ref}})]}. \quad (18)$$

The CPDF, $F_i$, monotonically increases from 0 to 1 and represents the probability of finding a phonon with a wavevector index smaller than $i$. To initialize the phonon wavevector, a random number $R_{\text{w}}$ between 0 and 1 is generated. If $F_{i-1} < R_{\text{w}} \leq F_i$ is satisfied, the phonon is assigned with the $i$th wavevector. Once the phonon wavevector is assigned, another CPDF for phonon branch can be calculated as

$$P_{i,j}^{\text{ph}} = \frac{\sum_{\beta=1}^{j} [n^0(\omega_{\beta i}, T_0) - n^0(\omega_{\beta i}, T_{\text{ref}})]}{\sum_{\beta=1}^{N_\beta} [n^0(\omega_{\beta i}, T_0) - n^0(\omega_{\beta i}, T_{\text{ref}})]}. \quad (19)$$

Another number $R_{\text{b}}$ between 0 and 1 is generated. If $P_{i-1,j} < R_{\text{b}} \leq P_{i,j}$ is satisfied, the phonon is assigned to the $j$th branch. As can be seen, the distribution function $n^0(\omega_{\beta q}, T_0) - n^0(\omega_{\beta q}, T_{\text{ref}})$ is positive if $T_0 > T_{\text{ref}}$ while negative if $T_0 < T_{\text{ref}}$. Therefore, a sign function $s = \pm 1$ is assigned to the phonon depending on the local temperature compared to the reference temperature.

Similarly, for electrons, two CPDFs for wavevector and band index sampling are calculated from the electronic band structure as

$$F_i^e = \frac{\sum_{k=1}^{i}\sum_{\alpha=1}^{N_\alpha}\left\{\frac{1}{\left[\exp\left(\frac{|\epsilon_{\alpha k} - \epsilon_F|}{k_B T_0}\right) + 1\right]} - \frac{1}{\left[\exp\left(\frac{|\epsilon_{\alpha k} - \epsilon_F|}{k_B T_{\text{ref}}}\right) + 1\right]}\right\}}{\sum_{k=1}^{N_k}\sum_{\alpha=1}^{N_\alpha}\left\{\frac{1}{\left[\exp\left(\frac{|\epsilon_{\alpha k} - \epsilon_F|}{k_B T_0}\right) + 1\right]} - \frac{1}{\left[\exp\left(\frac{|\epsilon_{\alpha k} - \epsilon_F|}{k_B T_{\text{ref}}}\right) + 1\right]}\right\}}, \qquad (20)$$

$$P_{i,j}^e = \frac{\sum_{\alpha=1}^{j}\left\{\frac{1}{\left[\exp\left(\frac{|\epsilon_{\alpha i} - \epsilon_F|}{k_B T_0}\right) + 1\right]} - \frac{1}{\left[\exp\left(\frac{|\epsilon_{\alpha i} - \epsilon_F|}{k_B T_{\text{ref}}}\right) + 1\right]}\right\}}{\sum_{\alpha=1}^{N_\alpha}\left\{\frac{1}{\left[\exp\left(\frac{|\epsilon_{\alpha i} - \epsilon_F|}{k_B T_0}\right) + 1\right]} - \frac{1}{\left[\exp\left(\frac{|\epsilon_{\alpha i} - \epsilon_F|}{k_B T_{\text{ref}}}\right) + 1\right]}\right\}}. \qquad (21)$$

Two random numbers are generated and compared to the CPDFs, such that the wavevector and band index can be assigned to the electron. If $T_0 > T_{\text{ref}}$, the deviational electron is assigned a positive sign. Otherwise, a negative sign is assigned.

After the wavevector and branch/band indices are assigned to the phonon/electron, the velocity and frequency/energy is automatically assigned according to the phonon dispersion and electron band structure obtained from DFT calculations.

c) Drift

In the drift step, all electrons and phonons travel ballistically. The position of a particle is updated by

$$\vec{r}(t + \Delta t) = \vec{r}(t) + \vec{v}\Delta t, \qquad (22)$$

where $\Delta t$ is the timestep.

d) Update cell temperature and energy

After the drift step, each cell energy can be calculated by summing the electron/phonon energies over all particles located in the cell as

$$\begin{cases} E^{\text{ph}} = \dfrac{1}{l_x l_y l_z} \sum_{j \in \text{cell}} \hbar \omega_j s_j \\ E^{\text{e}} = \dfrac{1}{l_x l_y l_z} \sum_{j \in \text{cell}} |\epsilon_j - \epsilon_F| s_j \end{cases}. \tag{23}$$

The cell energies obtained above are then plugged into Eq. (8) and (11) to determine the local electron and phonon temperatures of each cell.

e) Scattering

After computing the local temperature, the scattering probability is determined as

$$P = 1 - e^{-\frac{\Delta t}{\tau}}, \tag{24}$$

where the relaxation time $\tau$ is obtained from Eq. (4). We follow the steps given in the deviational MC scheme of phonon [50] and electron [51] transport. The implementations of electron and phonon scattering are the same. First, a pool of electrons/phonons to be scattered is determined by generating a random number and comparing it to the scattering probability given by Eq. (24) for each particle. Let $N^+$ be the number of electrons/phonons with positive sign and $N^-$ be the number of electrons/phonons with negative sign. Then, a total number of $N^+ + N^- - |N^+ - N^-|$ electrons/phonons from the pool are deleted, whereas the remaining electrons/phonons in the pool are resampled. The resampling distribution for electrons is $\frac{1}{\tau} \left| \dfrac{1}{\exp\left(\frac{|\epsilon_{\alpha k} - \epsilon_F|}{k_B T_e}\right)+1} - \dfrac{1}{\exp\left(\frac{|\epsilon_{\alpha k} - \epsilon_F|}{k_B T_{\text{ref}}}\right)+1} \right|$ and that for phonons is $\frac{1}{\tau} \left| \dfrac{1}{\exp\left(\frac{\hbar \omega_{\beta q}}{k_B T_{\text{ph}}}\right)-1} - \dfrac{1}{\exp\left(\frac{\hbar \omega_{\beta q}}{k_B T_{\text{ref}}}\right)-1} \right|$. Such implementation of the scattering process is guaranteed to satisfy the law of energy conservation [50, 51].

f) Heat flux and thermal conductivity

The electron and phonon heat flux ($\vec{q}_e$ and $\vec{q}_{ph}$) in each cell is calculated by summing energy flow contributed by all electrons/phonons in the cell

$$\begin{cases} \vec{q}_e = \dfrac{1}{l_x l_y l_z} \displaystyle\sum_{j \in \text{cell}} W|\epsilon_j - \epsilon_F|\vec{v}_j s_j \\ \vec{q}_{ph} = \dfrac{1}{l_x l_y l_z} \displaystyle\sum_{j \in \text{cell}} W\hbar\omega_j \vec{v}_j s_j \end{cases}, \quad (25)$$

from which the total heat flux can be easily calculated as $\vec{q} = \vec{q}_e + \vec{q}_{ph}$. In the 1D situation where a temperature gradient $\nabla T$ is imposed along a particular direction, the drift and scatter processes drive the system to a steady state in which a linear temperature profile is established. In this case, the thermal conductivity, $\kappa_z$, can be calculated using Fourier's law of heat conduction in 1D,

$$\kappa_z = -\frac{\bar{q}_z}{\nabla T}, \quad (26)$$

where z represents the direction of 1D heat flow and $\bar{q}_z$ is average of $q_z$ over all cells.

## 2.4 Density functional theory calculations

For simplicity, let $\lambda = \beta q$ denote a phonon mode. The ph-ph relaxation time due to three phonon scattering is given by the Fermi's golden rule as [6, 20]

$$\frac{1}{\tau_\lambda^{ph-ph}} = \frac{\pi\hbar}{16 N_\beta N_q} \sum_{\lambda_1 \lambda_2} |\Psi_{\lambda\lambda_1\lambda_2}|^2 \{(n_{\lambda_1}^0 + n_{\lambda_2}^0 + 1)\delta(\omega_\lambda - \omega_{\lambda_1} - \omega_{\lambda_2}) + (n_{\lambda_1}^0 - n_{\lambda_2}^0)[\delta(\omega_\lambda + \omega_{\lambda_1} - \omega_{\lambda_2}) - \delta(\omega_\lambda - \omega_{\lambda_1} + \omega_{\lambda_2})]\}, \quad (27)$$

where $\delta$ is the Dirac delta function. The ph-ph scattering matrix $\Psi_{\lambda\lambda_1\lambda_2}$ can be calculated from the third-order force constants [20, 52].

Let $\phi = \alpha k$ denote an electron state. Both the ph-e and e-ph relaxation time can be obtained from Fermi's golden rule as [6]

$$\frac{1}{\tau_\lambda^{ph-e}} = \frac{2\pi}{\hbar} \sum_{\phi_1 \phi_2} |\Phi_\lambda^{\phi_1 \phi_2}|^2 \{(f_{\phi_1}^0 - f_{\phi_2}^0)\delta(\epsilon_{\phi_1} - \epsilon_{\phi_2} + \hbar\omega_\lambda)]\}, \quad (28)$$

$$\frac{1}{\tau_\phi^{e-ph}} = \frac{2\pi}{\hbar} \sum_\lambda \sum_{\phi_1} \left|\Phi_\lambda^{\phi\phi_1}\right|^2 \{(n_\lambda^0 + f_{\phi_1}^0)\delta(\epsilon_\phi + \hbar\omega_\lambda - \epsilon_{\phi_1}) \tag{29}$$
$$+ (n_\lambda^0 + 1 - f_{\phi_1}^0)\delta(\epsilon_\phi - \hbar\omega_\lambda + \epsilon_{\phi_1})\},$$

where $\Phi_\lambda^{\phi\phi_1}$ is the e-ph scattering matrix. This matrix represents the scattering events where an electron at initial state $\phi$ is scattered into a new state $\phi_1$ by a phonon mode $\lambda$. From quantum mechanics, the e-ph scattering matrix can be calculated as [15]

$$\Phi_\lambda^{\phi\phi_1} = \sqrt{\frac{\hbar}{2\omega_\lambda}} \langle\psi_\phi|\partial U_\lambda|\psi_{\phi_1}\rangle, \tag{30}$$

where $\psi$ is the electron wavefunction and $\partial U_\lambda$ represents the derivative of potential with respect to the phonon displacement.

We use the standard DFT methods for calculations on both $\alpha$-U and Ag in this work. For Uranium compounds, it is known that the standard density function theory (DFT) method might be insufficient to describe the correlation effects of $5f$ electrons. More advanced methods including DFT+U [53] and dynamical mean-field theory (DMFT) [54] are required to properly model the electronic properties of U compounds and alloys such as $UO_2$ [55], UN [56], and U-Zr alloys [57]. However, for pure U, it has been shown that the standard DFT can well describe its structural, electronic [58, 59], phonon, and thermal properties [52, 60]. All DFT calculations are performed using the Quantum Espresso package [61]. For the pseudopotential, the exchange and correlation functional is treated by the Generalized Gradient Approximation (GGA) [62].

In the ph-ph relaxation time calculations, the 2[nd]-order force constants are obtained using the density functional perturbation theory [63]. The 3[rd]-order force constants are obtained using the finite displacement method where an auxiliary python module *thirdorder.py* in the ShengBTE [20] code is used. A $4 \times 4 \times 4$ supercell and $3 \times 3 \times 3$ nearest neighboring atoms are used in calculating the 3[rd]-order force constants. In the ph-e and e-ph relaxation time calculations, the e-ph scattering matrix is calculated using the Electron-Phonon Wannier (EPW) package [64]. The e-ph scattering matrix is first calculated on a $10 \times 10 \times 10$ electron k-point mesh and a $6 \times 6 \times 6$ phonon q-point mesh. Then, using the maximally localized Wannier functions basis [65], the matrix elements are interpolated to a denser $40 \times 40 \times 40$ k-point mesh and $30 \times 30 \times 30$ q-point mesh.

## 3 Results and Discussion

### 3.1 Electronic, phonon, and scattering properties

In this section, we present the results of electron band structure, phonon dispersion, and the electron and phonon relaxation times obtained from DFT.

We have tested the total energy of the $\alpha$-U unit cell as a function of different DFT parameters and plot the data in Figure 2. An electron wavefunction cutoff of 120 Ry, a charge density cutoff of 480 Ry, a gaussian smearing with 0.005 Ry spreading width, and a $10 \times 10 \times 10$ Monkhorst-Pack [66] k-point mesh is found to provide sufficiently converged unit cell energy. Using the above parameters, the structure optimization is performed where both the unit cell shape and internal atomic coordinates are allowed to relax. The initial lattice configuration is taken from our previous study on the electron band structure and phonon dispersion of $\alpha$-U using the Vienna ab initio simulation package (VASP) [67], where the lattice parameters are $a = 2.817$ Å, $b = 5.867$ Å, and $c = 4.875$ Å. Full relaxation is performed until the average force acting on the atom is less than $1 \times 10^{-5}$ Ry/Bohr, and the energy difference between two consecutive relaxed configurations is less than $1 \times 10^{-6}$ Ry. The relaxed lattice parameters are $a = 2.817$ Å, $b = 5.867$ Å, and $c = 4.875$ Å.

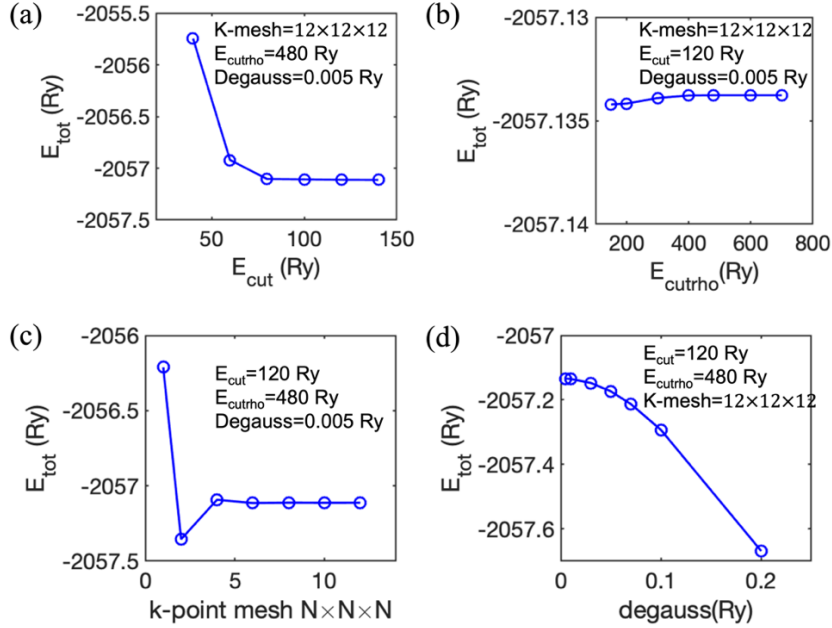

Figure 2. Convergence tests of total energy ($E_{tot}$) on the (a) energy cutoff for electron wavefunctions ($E_{cut}$) and (b) k-point mesh, (c) gaussian spreading of electron occupation function (degauss), and (d) energy cutoff for charge density ($E_{cutrho}$). In each test, the other three parameters are fixed at values listed in the figure. For the DFT parameters, large energy cutoff is desirable to include enough number of plane-wave basis functions, numbers of k-points should be large to ensure the accuracy of integration over the BZ, and the gaussian spreading should be as small as to limit the discrepancy between the smooth occupation function and the discontinuous Fermi-Dirac function. From all four figures, we extract the threshold values of parameter as: $E_{cut} \geq 80$ Ry, $E_{cutrho} \geq 400$ Ry, k-mesh$\geq 6 \times 6 \times 6$, and degauss $\leq 0.1$ Ry.

After obtaining the equilibrium lattice configuration, the phonon dispersion and electronic band structure of $\alpha$-U is calculated and presented in Figure 3. Both band structure and DOS of phonons and electrons agree well with our previous calculations obtained using VASP [52], as well as other experimental and simulation results. By comparing Figure 3(c) and Figure 3(d), it is clear that the phonon group velocity ranges from 10 to 1000 m/s while the electron velocity ranges from 1 to 1000 km/s. The electron velocity is generally several orders of magnitude larger than the phonon velocity. This indicates much faster diffusion of electrons than phonons in the thermal transport process.

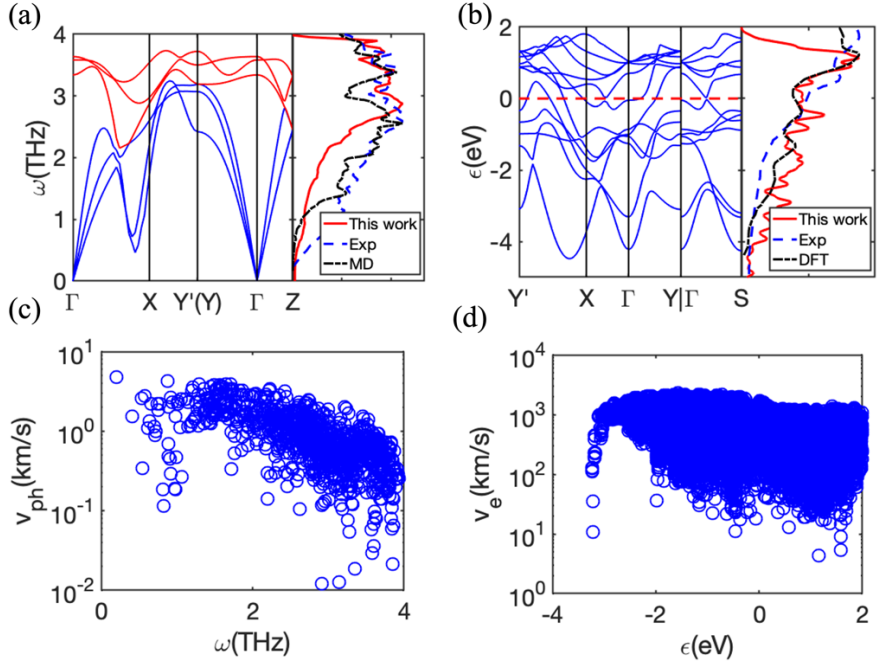

Figure 3. (a) phonon dispersion and density of states (DOS), (b) electronic band structure and DOS, (c) phonon group velocity, and (d) electronic velocity of $\alpha$-U. In Figure (a), the red lines represent optical modes while the blue lines represent the acoustic modes. The phonon DOS is compared with data obtained from neutron scattering experiments (Exp) at 50K [68] and MD simulation [69]. The electron DOS data is compared with DFT calculations [57] and X-ray photoemission spectroscopy measurements [70].

Using the obtained phonon dispersion and electron band structure, the ph-ph, ph-e, and e-ph scattering of $\alpha$-U is calculated and presented in Figure 4. In Figure 4(a), the ph-ph relaxation time at 100K, 300K, and 500K are plotted which shows a consistent decrease in the relaxation time as temperature increases. This is due to more frequent ph-ph scattering events at higher temperature. In Figure 4(b), we show the Eliashberg function and the integrated e-ph coupling constant as a function of phonon frequency. Note that the Eliashberg function $\alpha^2 F(\omega) \approx 0$ for phonons with frequency lower than around 1.5 THz. This signifies that low-frequency acoustic phonons do not readily exchange energy with electrons. The largest contribution to the ph-e coupling arises from the high frequency acoustic phonons, made evident by the peak in the Eliashberg function around 2.1 THz. Our calculated e-ph coupling constant of $\lambda = 0.52$ compares well with $\lambda = 0.44$ from another DFT calculation [71]. As our k-point and q-point mesh ($30 \times 30 \times 30$) is significantly denser than the ones ($8 \times 6 \times 6$) used in their work, we believe our calculated $\lambda$ is more accurate. As seen in Figure 4(c) and (d), the e-ph relaxation time is one order of magnitude smaller than the ph-e relaxation time. This indicates that e-ph scatterings drive electrons toward thermal equilibrium at a much faster rate than that of phonons.

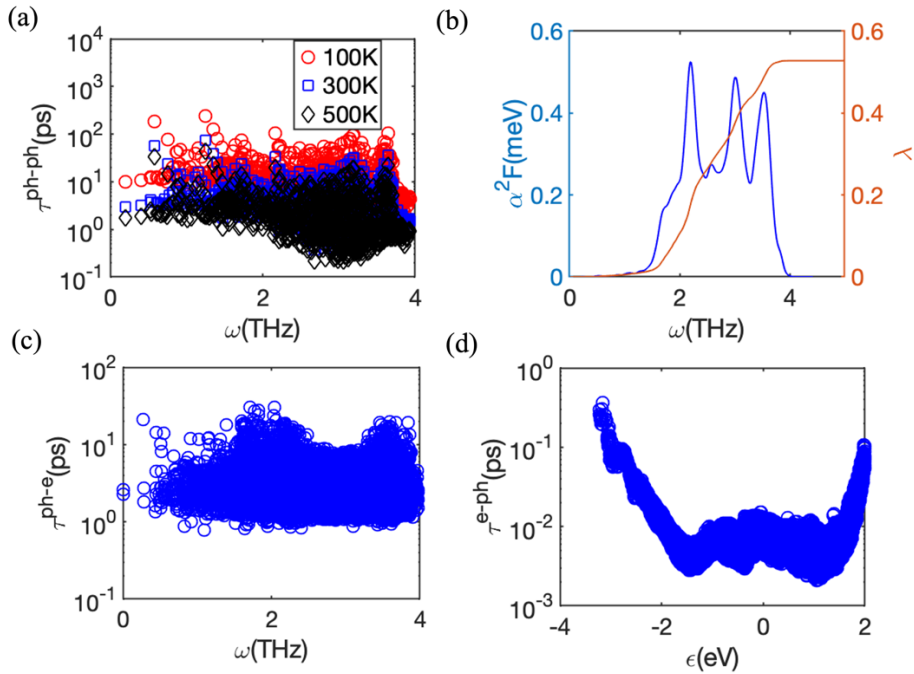

Figure 4. (a) Ph-ph relaxation time, (b) Eliashberg function and e-ph coupling constant, (c) ph-e relaxation time, and (d) e-ph relaxation time of $\alpha$-U. In all calculations, a $30 \times 30 \times 30$ grid is used for both electron and phonon mesh.

Our convergence tests of the Ag unit cell total energy with respect to DFT parameters are shown in Figure 5. An electron wavefunction cutoff of 60 Ry, a charge density cutoff of 240 Ry, a gaussian smearing with 0.025 Ry spreading width, and a $12 \times 12 \times 12$ Monkhorst-Pack k-point mesh is chosen. After relaxation, the equilibrium lattice parameter is $a = 2.817$ Å.

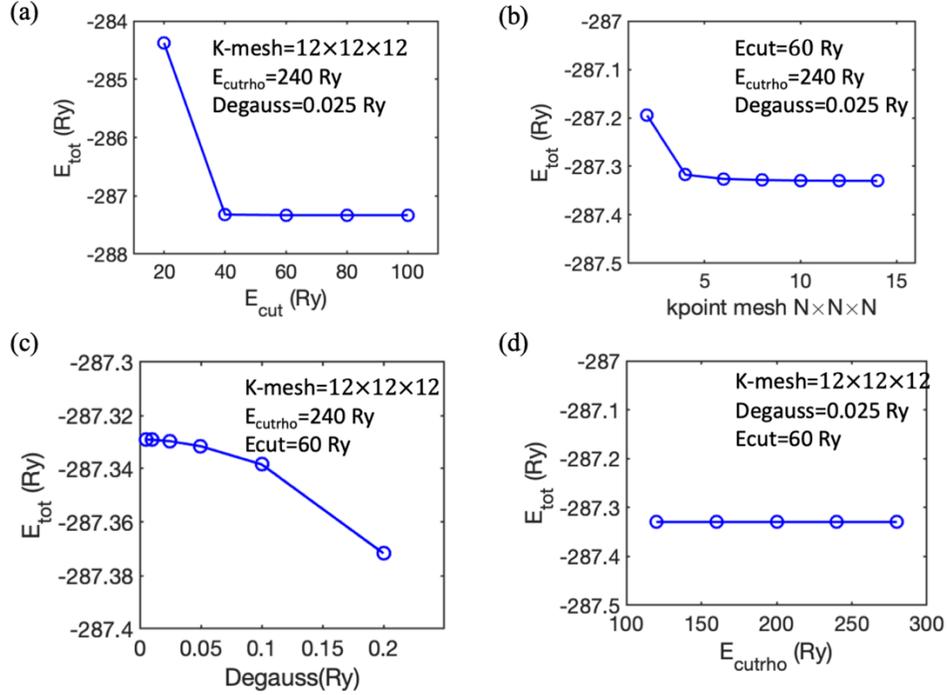

Figure 5. Convergence tests of DFT calculations of Ag. All labels are the same as Figure 2. From all four figures, the threshold values of parameters are: $E_{cut} \geq 60$ Ry, $E_{cutrho} \geq 120$ Ry k-mesh$\geq 6 \times 6 \times 6$ Ry, and degauss $\leq 0.1$ Ry.

The phonon and electronic band structure of Ag is plotted in Figure 6. As can be seen in Figure 6(a) and (c), the phonon group velocity is inversely related to its frequency, with the lowest ZA branch phonons having the largest velocity. In Figure 3(b), the electron DOS obtained from our calculations, which agrees well with previous works [72, 73], predicts the largest population of electrons to be located near a band energy of -5 eV. However, the electrons with energies near 0 eV, that of the Fermi level, transport the most thermal energy. Despite the relatively small DOS around Fermi level, Ag still has a high thermal conductivity due to its large population of free electrons. Comparison of Figure 6(c) and (d), shows the electron velocity to be approximately three orders of magnitude larger than that of phonons.

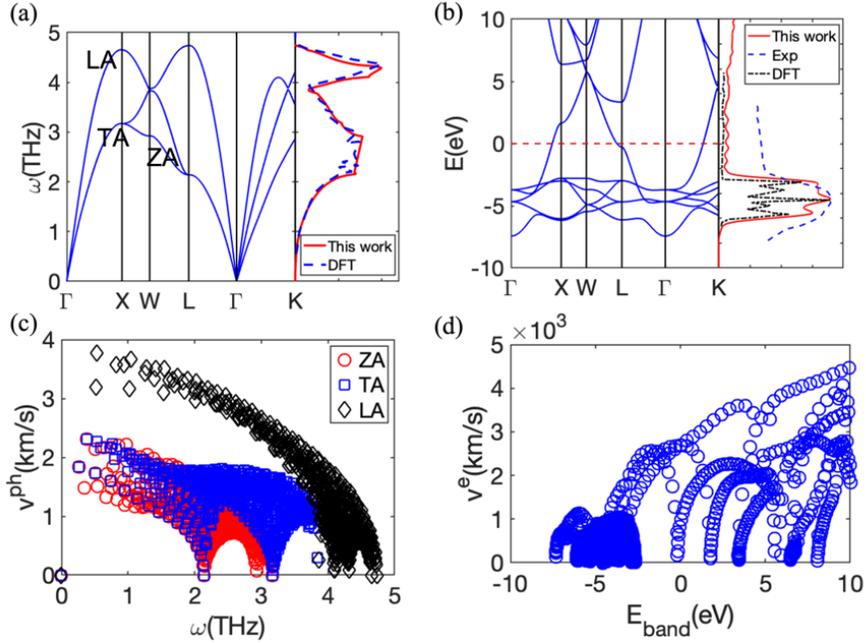

Figure 6. (a) phonon dispersion and density of states (DOS), (b) electronic band structure and DOS, (c) phonon group velocity, and (d) electronic velocity of Ag. In Figure (a), the phonon DOS is compared with data obtained from another DFT calculation [74]. In Figure (b), the electron DOS is compared with data obtained from experiments [72] and DFT calculations [73].

Our calculated relaxation times and the e-ph coupling constant for Ag are presented in Figure 7. In Figure 7(a), the ph-ph relaxation times decrease with increasing temperature. Furthermore, the peak in relaxation times near 3.3 THz predicts that phonons near this frequency are less likely to participate in ph-ph scattering. In Figure 7(b), the two peaks of the Eliashberg function at 3 and 4.3 THz coincide with the peaks in the phonon DOS in Figure 6(a). These phonons are higher frequency modes, mostly in the LA branch. Therefore, due to their strong e-ph coupling strength, the ph-e relaxation times of LA phonons are smaller than those of the phonons in ZA and TA branch. Inspection of Figure 6(c) confirms the ZA and TA phonons to have larger ph-e relaxation times than the LA phonons. Lastly, by comparing Figure 7(d) with Figure 7(a) and (c), the e-ph relaxation time is approximately three orders of magnitude smaller than the phonon relaxation times.

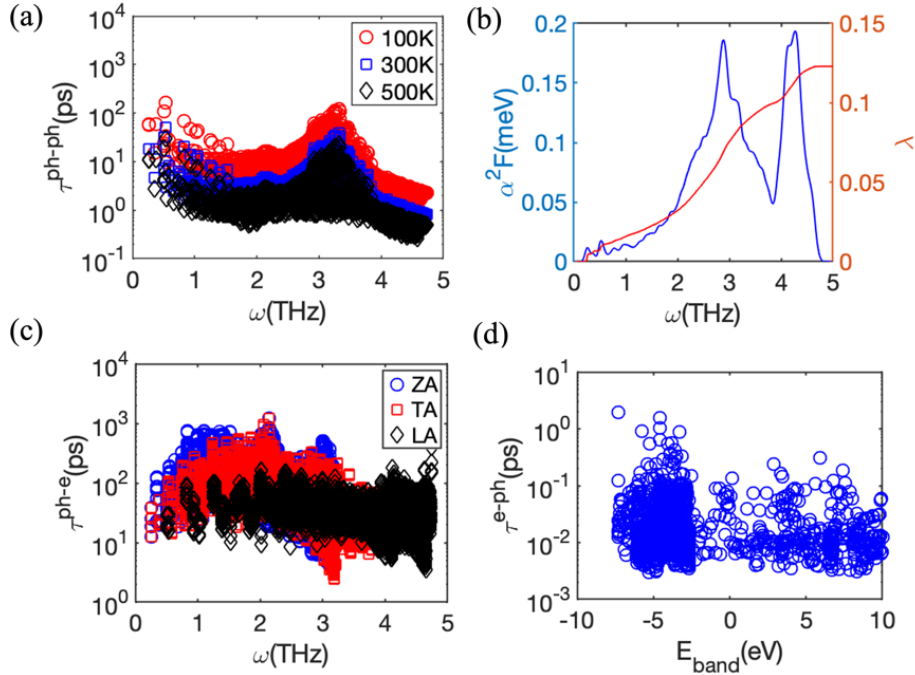

Figure 7. (a) Ph-ph relaxation time, (b) Eliashberg function and e-ph coupling constant, (c) ph-e relaxation time, and (d) e-ph relaxation time of Ag.

# 4 MC simulation of coupled electron-phonon thermal transport

## 4.1 Verification and validation

In this section, the MC algorithm developed is verified by two case studies. The phonon BTE algorithm is verified by MC simulation of 1D phonon transport in $\alpha$-U, while the electron BTE algorithm is verified by 2D electron thermal transport in Ag.

In 1D phonon transport, there are two limiting cases where an analytical solution of the phonon BTE can be derived – the diffusive and ballistic phonon transport limit. When the characteristic size of the system is much larger than the phonon mean free path, frequent scattering events restore local thermal equilibrium and the transport is diffusive; when the system size is comparable to or smaller than the phonon mean free path, scattering events rarely occur and the transport is mainly ballistic.

In diffusive phonon transport, an analytical solution of the 1D heat transfer equation can be obtained with Laplace's transform as [75]

$$\frac{T(z,t) - T(L,t)}{T(0,t) - T(L,t)} = \left[\text{erfc}\left(\frac{z}{2\sqrt{\alpha t}}\right) - \text{erfc}\left(\frac{2L-z}{2\sqrt{\alpha t}}\right) + \text{erfc}\left(\frac{2L+z}{2\sqrt{\alpha t}}\right)\right], \quad (31)$$

where $\text{erfc}(s)$ is the complementary error function of $s$, $z$ is the distance from the hot cell (see Figure 8(a)), $\alpha$ is the thermal diffusivity, and $L = N_z l_z$ is the total length along the z-direction. The thermal diffusivity can be calculated as $\alpha = \frac{\kappa}{\rho C_p}$, where $\kappa$ is thermal conductivity, $\rho$ is density, and $C_p$ is the specific heat capacity. For $\alpha$-U, an experimental value of $\alpha = 0.12\ cm^{-2}/s$ is used [76]. A schematic of the 1D simulation domain is shown in Figure 8(a), and the corresponding parameters are $L = 1500$ nm, $N_x = N_y = 1$, $N_z = 30$, $l_x = l_y = 200$ nm, $l_z = 50$ nm, $T_h = 310$ K, $T_c = 290$ K. The temperature profiles obtained from our MC simulations and shown in Figure 8(b) show excellent agreement with the analytical solution given by Eq. (31). At $t = 20$ ns, the system has already achieved the steady state where a linear temperature profile has been established.

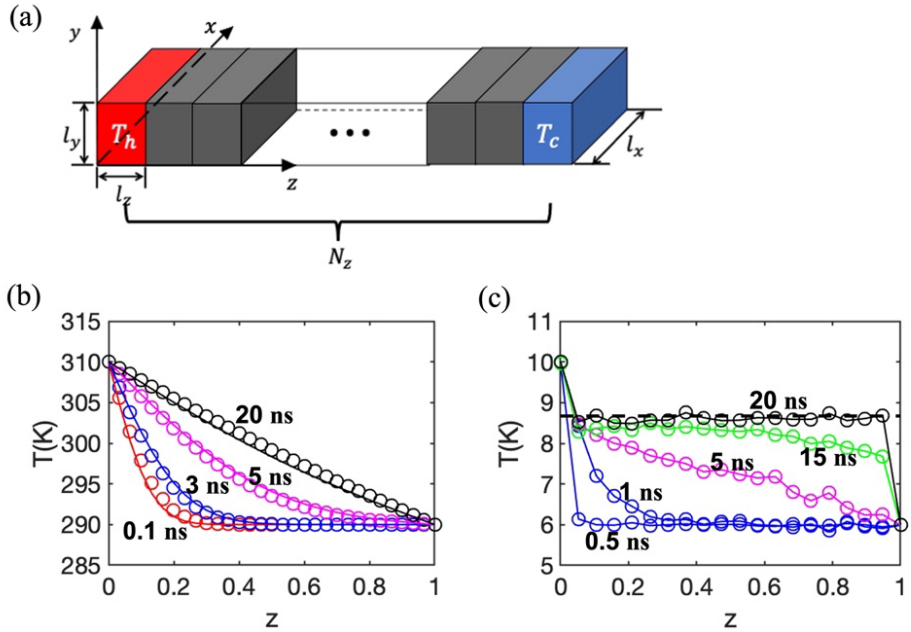

Figure 8. (a) Schematic of MC simulation domain of 1D phonon transport in $\alpha$-U. The dimension is discretized only along z-direction. Although phonons can travel in any direction in 3D, the imposed temperature gradient along z-direction enables 1D heat transfer. (b) Transient temperature profile of diffusive phonon transport. (c) Transient temperature profile of ballistic phonon transport. In both (b) and (c), the MC simulation data is represented by scattered points in comparison with the analytical solution represented by solid lines. Initially all the interior cells are assigned with temperature $T = T_c$. In MC simulations, the phonon scattering/diffusion change the cell temperature. The temperature values are obtained from five independent MC simulations with different initial conditions including the position, wavevector and branch index. In each MC run, the temperature at a specific simulation time is averaged over a 20 ps time interval.

In ballistic phonon transport, the steady state temperature profile is given by the Stefan-Boltzmann law [27]

$$T(z) = \left(\frac{T_h^4 + T_c^4}{2}\right)^{\frac{1}{4}}. \tag{32}$$

The simulation parameters are $L = 1500$ nm, $N_x = N_y = 1$, $N_z = 20$, $l_x = l_y = 200$ nm, $l_z = 50$ nm, $T_h = 10$ K, and $T_c = 6$ K. In Figure 8(c), a temperature discontinuity between two end cells and the first interior cell appears due to spatial discretization [52, 27]. At $t = 20$ ns, the steady-state temperature profile obtained from MC simulation agree well with the analytical solution $T(z) = 8.67$ K given by Eq. (32).

Next, we continue to validate our MC method by simulating 2D electron thermal transport in a thin Ag slab. The slab is bounded by two walls at $y = 0$ and $y = d$ where diffusive reflecting boundary conditions are imposed. The diffusive boundary is implemented by generating an electron which has the same energy and velocity magnitude with the incident electron, in a random traveling direction. Boundaries in the x-direction are treated as periodic and the temperature gradient is along the z-direction. The simulation parameters are $L_x = 200$ nm, $l_z = 50$ nm, $N_x = 1$, $N_y = 25$, $N_z = 20$, $T_h = 310$ K, and $T_c = 290$ K. After reaching steady-state, the temperature and heat flux of each cell are averaged over 200 timesteps. The steady state temperature profile of discretized cells in the yz plane are shown in Figure 9(a) where a linear temperature gradient along the z-direction is established. Along the y-direction, the temperature of cells close to the diffusive wall is slightly lower than the interior cells because of electron-boundary scattering. The enhancement of boundary scattering is more clearly seen in Figure 9(b), where the heat flux assumes a parabolic shape in the y-direction with a maximum in the middle. In Figure 9(c), we show the spatial variant temperature along the z-direction at $y = 0.5$. The initial temperature of all cells was set as 300 K. As the simulation proceeds to the final timestep at $t = 40$ ps, a linear temperature profile has been established. To calculate the effective thermal conductivity, we plot the time dependent heat flux along the z-direction in Figure 9(d). Near $t = 5$ ps, the stabilized heat flux signifies that a steady-state has been reached which corresponds to a linear temperature profile. The electronic thermal conductivity $\kappa^e = 431.3$ W/m/K is obtained from the heat flux data collected after 10 ps. This value is in reasonable agreement with 436 W/m/K from experiment [8, 11] and 450.86 W/m/K [6] from DFT calculations.

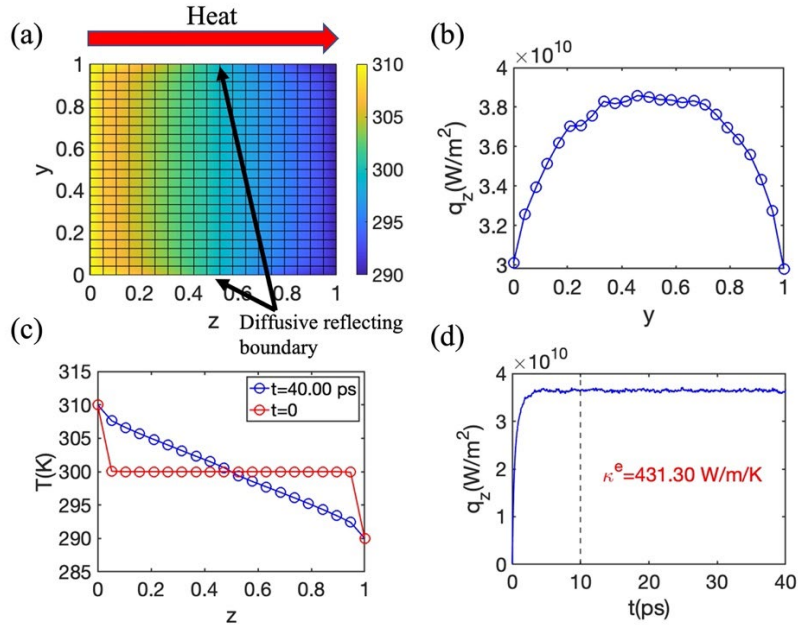

Figure 9. MC simulation of Ag electron transport in a thin slab. (a) Temperature profile, (b) heat flux along y-direction, (c) initial and steady-state temperature profile along z-direction, and (d) effective electronic thermal conductivity of the slab.

Next, we consider the effects of varying slab thickness. Here, the analytical solution of the BTE can be obtained from the classical Fuchs-Sondheimer (F-S) model [50, 77]. In Figure 10(a), the heat flux profile obtained from our MC simulations agree well with the F-S model, with small fluctuations due to the stochastic nature of the MC method. With decreased thickness, the heat flux near the boundary is reduced due to enhanced electron-boundary scattering. Eventually, an almost uniform heat flux profile will appear as the thickness approaches almost zero. Since only electron-boundary scattering is present, the electron transport is purely ballistic and a similar temperature profile to the ballistic phonon transport will be present, as shown in Figure 8(c). The effective thermal conductivity of the two slabs with different thickness are shown in Figure 10(b), where the good agreement between conductivity by MC simulation and F-S model further validates our MC algorithm of the electron thermal transport.

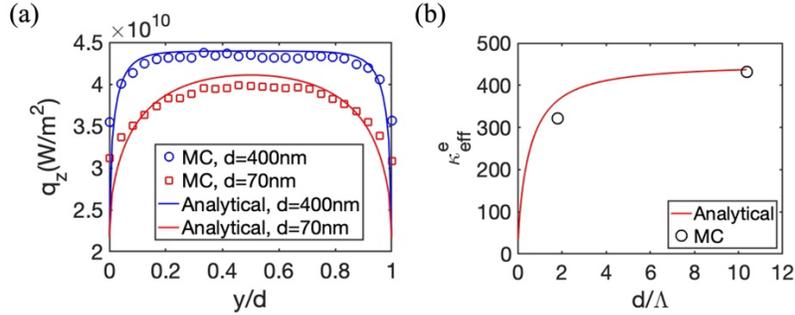

Figure 10. (a) Heat flux and (b) effective thermal conductivity of Ag thin slab obtained from the MC simulation and FS model.

## 4.2 Coupled e-ph transport

In this section, we combine the electron and phonon transport algorithms validated in the previous section and implement the e-ph scattering algorithm to study coupled e-ph transport in Ag and $\alpha$-U. First, we performed e-ph coupled thermal transport of $\alpha$-U at 300 K. Both electrons and phonons are allowed to drift and scatter in the domain. The simulation parameters are: $l_x = 500$ nm, $l_y = 20$ nm, $l_z = 50$ nm, $N_x = 1$, $N_y = 25$, $N_z = 20$, $T_h = 310$ K, and $T_c = 290$ K.

Electron temperature profiles are shown in Figure 11(a), where it requires 20 ps for electrons to reach steady state, signified by a linear temperature profile. Similarly, in Figure 11(b), phonons require 10 ns to achieve steady state. Note that the electron temperature referred to here is the thermodynamic temperature of the electron ensemble, which is different from the previous work that adopts a kinetic definition for the electron temperature in the MC simulation [30]. The kinetic expression used in the previous work overestimates the average energy and temperature of electron ensembles. The time for electrons and phonons to reach equilibrium is decided by the electron/phonon velocity and the time scale of scattering events, namely, their relaxation times. Because the electron velocity is much larger than the phonon velocity, and the electron relaxation time is much smaller than the phonon relaxation time, the equilibration time of phonons is much longer than that of electrons as we see in Figure 11(a) and (b). In Figure 11(c), we show the spatial variance of the electron and phonon heat flux. In the case of a linear temperature profile, the heat flux should be a constant over the domain which is verified by our MC simulation aside from stochastic fluctuations. We also observe the electron heat flux to be higher than the phonon heat flux which indicates that electrons dominate thermal transport in $\alpha$-U. In Figure 11(d), we plot the time dependent electron and phonon heat flux. Consistent with the temperature profile, the electron heat flux reaches equilibrium before 1 ns, whereas the phonon heat flux reaches steady state at approximately $t = 10$ ns. The electronic thermal conductivity $\kappa^e = 14.26$ W/m/K and phonon thermal conductivity $\kappa^{ph} = 8.34$ W/m/K is obtained from the heat flux data. The total thermal conductivity of 22.6 W/m/K falls within the range of 21-26 W/m/K from experimental measurements [76, 78, 79].

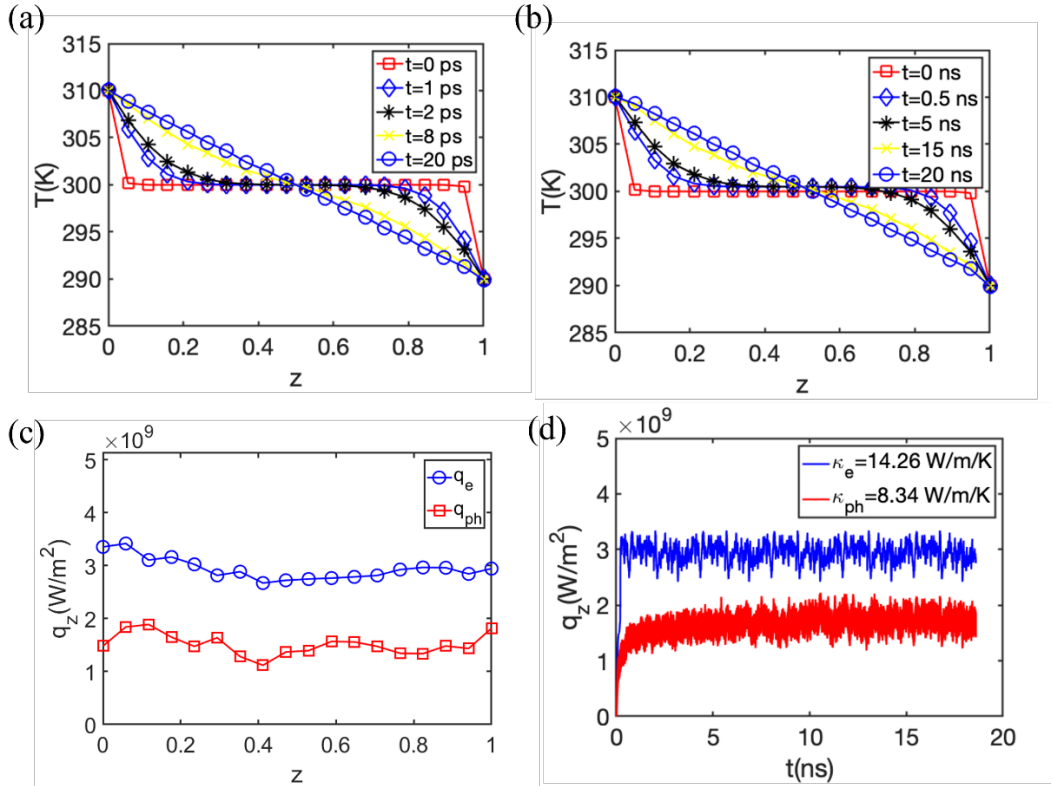

Figure 11. (a) Electron temperature and (b) phonon temperature profiles, (c) heat flux profile along z-direction, and (d) time-dependent heat flux of MC simulation of coupled e-ph transport in $\alpha$-U.

Next, we perform MC simulation of coupled e-ph thermal transport in Ag while varying temperature from 100 K to 1000 K. At each temperature, five different MC simulations with different initial electron/phonon ensembles (generated from different random number seeds) are conducted. Then, the thermal conductivity obtained from these simulations are averaged and presented in Figure 12(a). The phonon thermal conductivity from our MC simulations compares well with results from another DFT study that integrates conductivity from all phonon modes over the BZ [80]. To examine the effect of e-ph coupling on thermal conductivity of Ag, we performed another set of MC simulations where only ph-ph scatterings are considered. In Figure 12(a), the conductivity obtained from MC simulations with only ph-ph scatterings is only slightly higher than the values obtained from MC simulations with both ph-ph and ph-e scatterings. We conclude that phonon transport in Ag is dominated by ph-ph scattering. Furthermore, the contribution from ph-e scattering to the phonon thermal conductivity becomes smaller as temperature increases. At around 800 K, the effect of ph-e scattering on the phonon thermal conductivity is almost negligible. To quantitatively investigate the effect of ph-e scattering on the relaxation time of individual phonon modes, we define a ratio $r = 1 - \frac{\tau^{ph-e\&ph-ph}}{\tau^{ph-ph}}$ to represent the effect of ph-e scattering on the total phonon relaxation time. The ratio $r$ at 300 K of all phonon modes are plotted in Figure 12(b). As can be seen, the largest $r \approx 0.35$ appears in phonons with frequency around 3 THz. This agrees with the location of ph-e scattering strength peak shown in Figure 7(b). However, as shown in the phonon dispersion in Figure 6(a), most of these modes are close to the Brillouin zone edge with low group velocity and thus contribute little to the conductivity. The major contribution to the conductivity comes from the modes in the low-frequency region where $r < 0.2$. Overall, most phonons modes have a ratio of $r < 0.2$. As a result, the conductivity with both ph-e and ph-ph interactions does not differ much from the one that includes only ph-ph interaction, as shown in Figure 12(a). Inspection of Figure 12(c) shows the electronic thermal conductivity from our MC simulations to be slightly lower than the values obtained from an analytical model [81]. This is because the analytical model assumes isotropic and mode-independent e-ph scattering, thus providing an upper bound to the electron thermal conductivity [48]. Therefore, our conductivity values are reasonable. In Figure 12(d), we plot the percentage contribution to the total thermal conductivity from both phonons and electrons. At all temperatures, phonons contribute less than 2% to the total thermal conductivity. Thus, it is reasonable to only consider thermal transport by electrons in Ag.

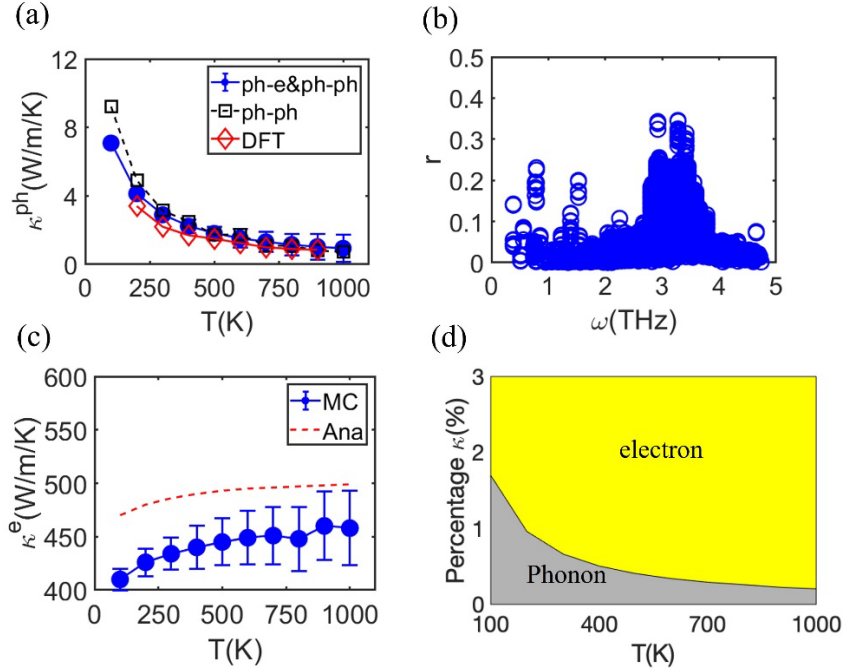

Figure 12. MC simulation of coupled e-ph thermal transport in Ag. (a) Phonon thermal conductivity as a function of temperature obtained with only ph-ph scattering and with both ph-e and ph-ph scattering included. (b) Reduction of the total phonon relaxation time of each phonon mode due to ph-e coupling. (c) Electron thermal conductivity as a function of temperature compared to an analytical mode [81]. (d) Percentage contribution to the total thermal conductivity from electrons and phonons.

Next, we investigated the anisotropic thermal conductivity of $\alpha$-U by MC simulations. In Figure 13(a), we plot the temperature dependent phonon thermal conductivity with and without ph-e scattering as a function of temperature. The reduction of phonon thermal conductivity along all three directions due to ph-e scattering is small at all temperature. This is consistent with our previous study on thermal conductivity of $\alpha$-U that uses combined DFT and empirical models of phonon relaxation times [52]. The mode-dependent reduction of phonon relaxation time brought by ph-e scattering is shown in Figure 13(b). For all phonon modes, the ph-e reduction ratio $r <$ 0.2 which is smaller than the reduction ratio of some modes in Ag, as can be seen in Figure 12(b). This indicates that ph-e scattering has a weaker impact on the phonon thermal conductivity of $\alpha$-U phonon thermal conductivity than in Ag. In Figure 13(c), we show the anisotropic electron thermal conductivity. The thermal conductivity along the [001] direction is slightly smaller than the conductivity along the [100] and [010] directions due to smaller electron velocities along the [001] direction. The percentage contribution from electrons and phonons to the total thermal conductivity along the [100] direction is shown in Figure 13(d). Note that the contribution from phonons to the total thermal conductivity decreases as temperature increases. The phonons in $\alpha$-U contribute less than 10% to the total conductivity at $T > 400$ K whereas they contribute as large as 60% to the total conductivity at 100 K. This is different from Ag where the phonons have negligible contribution to the conductivity at all temperatures. Therefore, for $\alpha$-U, both electron and phonon transport must be considered to study thermal problems at relatively low temperature.

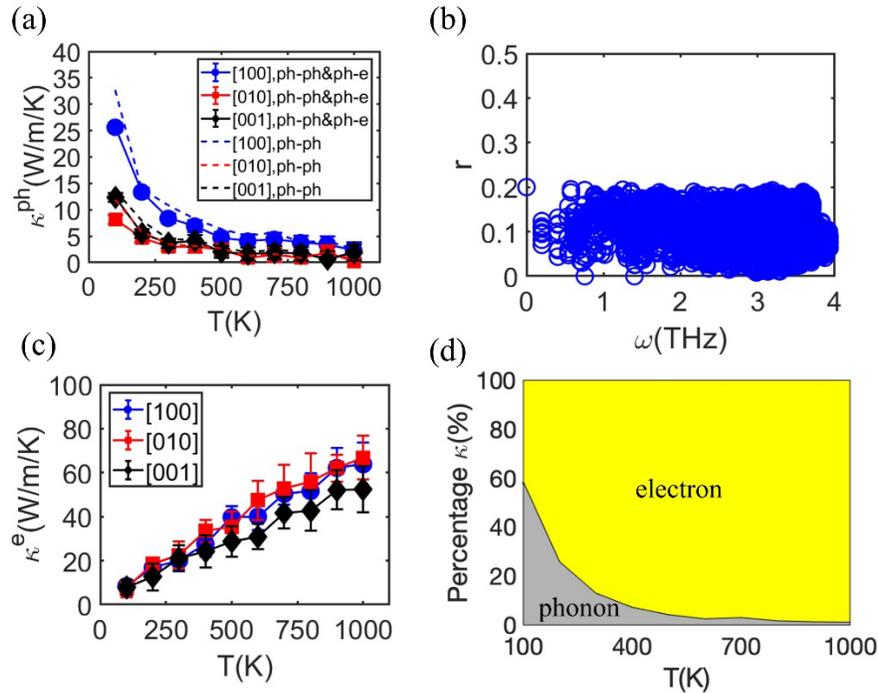

Figure 13. MC simulation of coupled e-ph thermal transport in $\alpha$-U. (a) Phonon thermal conductivity along three lattice directions obtained with only ph-ph scattering and with both ph-ph and ph-e scattering. (b) Reduction ratio $r$ of phonon relaxation time due to ph-e scattering. (c) Electron thermal conductivity along three lattice directions. (d) Percentage contribution to the total thermal conductivity from electron and phonon.

To further investigate the characteristics of coupled e-ph transport in $\alpha$-U and Ag, we calculate the Lorenz ratio $L = \frac{\kappa^e}{\sigma T}$ where $\sigma$ is the electronic conductivity. For simplicity, the electrical conductivity values are taken from our previous study on $\alpha$-U [52] that uses the Drude theory. In Figure 14(a), we plot the anisotropic $L$ of $\alpha$-U and isotropic $L$ of Ag as a function of temperature. At high temperature, $T > 800$ K, $L$ is in decent agreement with the Sommerfeld value. However, at low and intermediate temperatures, for instance $T = 100\ K$ and $T = 400 - 700$ K, $L$ can differ from the Sommerfeld value by around 20%. In contrast, another study used empirical expressions of relaxation times to calculate conductivity of $\alpha$-U and shows that the $L$ at all temperature only differs from the Sommerfeld value by 4% at most [52]. We believe our MC model which uses all DFT data of electron and phonon relaxation time provides a better estimate of the Lorenz ratio. Hence, we conclude that the widely employed Widemann-Franz law leads to inaccurate electronic thermal conductivity of $\alpha$-U and Ag at low to intermediate temperatures. Accurate calculation of thermal conductivity of metals requires full electron band structure, phonon dispersion, and mode-dependent e-ph relaxation time.

The electron and phonon mean free path (MFP) provides information on the size effect of thermal conductivity, which, in the case of Ag, is important for the nanoelectronic device applications. Therefore, we extract the cumulative electron and phonon thermal conductivity of $\alpha$-U and Ag from our MC simulations, which is plotted in Figure 14(b-d). In Figure 14(b), the majority of $\alpha$-U electron thermal conductivity comes from the electrons with thermal MFP smaller than 10 nm. In Ag, the electron MFP is considerably larger compared to the electron MFP of $\alpha$-U. This gives rise to a much large electron thermal conductivity of Ag than that of $\alpha$-U. In Ag, electrons with MFP between 10 and 100 nm make the most contribution to the electron thermal conductivity. In Figure 14(c) and (d), the phonon MFP decreases with increasing temperature due to enhanced ph-ph scattering at high temperature. At different temperature, the phonon MFP in $\alpha$-U and Ag are close to each other. Therefore, the phonon thermal conductivities in these two materials are close despite large difference in their electron thermal conductivity. By comparing the electron and phonon MFP of Ag, it can be seen that the electron MFP is larger than the phonon MFP. This indicates that the Ag electronic thermal conductivity exhibits a stronger size effect than the phonon thermal conductivity in nanostructures. Whereas for $\alpha$-U, the electronic and phonon thermal conductivity exhibit similar size effect.

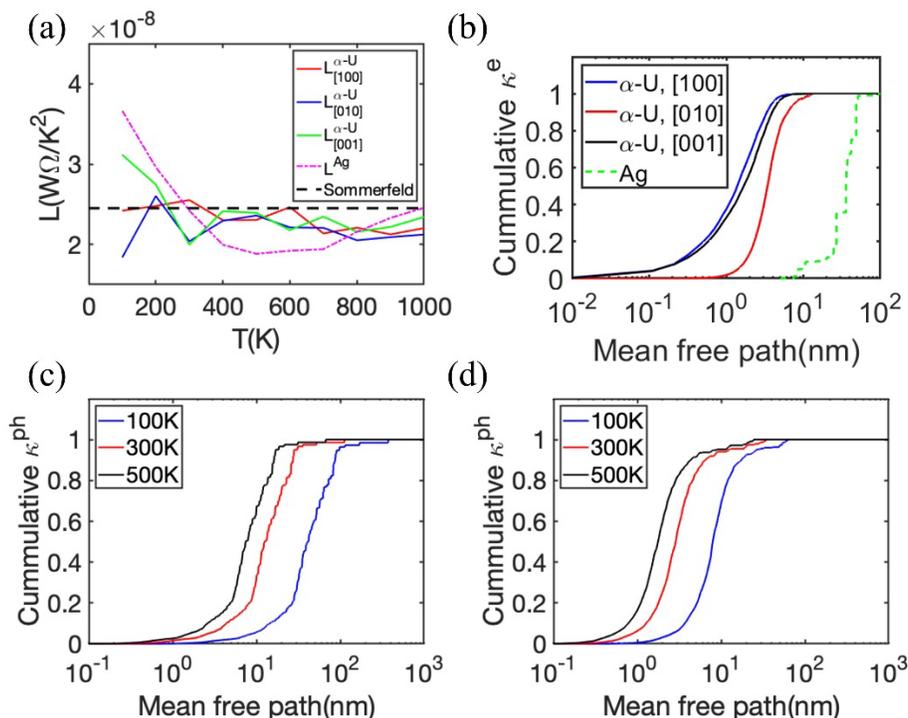

Figure 14. (a) Lorenz ration of $\alpha$-U and Ag obtained from MC simulations. The dash line represents the Sommerfeld value $2.45 \times 10^8$ W$\Omega$/K$^2$. (b) Cumulative electron thermal conductivity of $\alpha$-U and Ag. Cumulative phonon thermal conductivity of (c) $\alpha$-U and (d) Ag at three different temperatures.

## 5 Concluding remarks

We highlight several main contributions of this work. First, a MC approach for solving the coupled e-ph BTE is developed where electron band structure and phonon dispersion in the full BZ as well as the scattering among electrons and phonons are considered. This is a big improvement over majority of previous works that adopt overly simplified electron band, phonon dispersion, or scattering models [28, 29, 30]. Secondly, our approach enables study of 3D thermal transport in a device with complex geometry using electron and phonon properties obtained from first principles calculations. It can treat the details of scattering events at the microscopic scale and be applied to study anisotropic energy transport at the macroscopic scale. Understanding electron and phonon dynamics with microscopic detail can facilitate the development of novel electronics, energy devices, and spectroscopy techniques. Thirdly, the e-ph and ph-e scattering events are explicitly formulated via DFT into our MC scheme, as opposed to previous thermal transport models where e-ph scattering are ignored in phonon transport [10, 11] or empirical formulas are used [29, 30]. As the e-ph interaction is important in determining the intrinsic behavior of the electrical conductivity while causing scattering and joule dissipation in electronic devices [31, 32, 33, 34], our MC approach provides a strong tool to study thermal transport in these electronic devices. The developed methodology in this work can tackle fundamental thermal problems in solid-state physics and contribute to developing novel technological applications.

In this work, thermal transport in $\alpha$-U and Ag is investigated using a MC approach to solve the coupled e-ph BTE. The proposed MC scheme accounts for anisotropic phonon dispersion and electronic band structure in the full BZ. The ph-e relaxation time of every electron band and phonon mode are obtained from first principle DFT calculations. The developed MC scheme is verified by simulating room-temperature, 1D phonon thermal transport in $\alpha$-U and 2D electron thermal transport in Ag. Results obtained from our Monte Carlo method are in very good agreement with analytical solutions.

From the DFT calculations, we found that the electron relaxation time is orders of magnitude smaller than the phonon relaxation time. This is indicative of more frequent scattering events for electrons, which drives their population towards thermal equilibrium at a faster rate when compared to phonons. Regarding the phonon relaxation times, high frequency acoustic phonons were found to have the strongest e-ph coupling strength among all phonon modes, thus having the smallest relaxation time. From the MC simulations of coupled e-ph thermal transport, the ph-e scattering is found to have negligible influence on the phonon thermal conductivity. This is further studied by mode-dependent ph-e relaxation time analysis which shows that the ph-e interaction contributes less to the total relaxation time than the ph-ph interaction. Comparison between the electronic and phonon thermal conductivity shows that the electrons dominate thermal transport at high temperature in both $\alpha$-U and Ag. However, the phonon contribution to the thermal conductivity of $\alpha$-U is significant at low temperature, which can be as large as 60%. Furthermore, the Lorenz ratio obtained from our MC simulation deviates from the Sommerfeld value at low to intermediate temperatures. Therefore, the widely used Wiedemann-Franz law for metals can lead to inaccurate electronic thermal conductivity of metals. By analyzing the cumulative conductivity as a function of MFP, we find that the Ag electron thermal conductivity has a stronger size effect than the phonon thermal conductivity.

The developed MC approach can be extended to a larger time and spatial scale, where thermal transport in a device configuration can be simulated. Exotic materials such as 2D semiconductors, 1D van der Waals quantum materials, relaxor ferroelectrics, and topologic insulators have been tentatively applied to multijunction devices together with metals, traditional semiconductors, and insulators. They are supposed to overcome Boltzmann's tyranny due to dissipation, especially determinant in device behavior constraint by miniaturization. Many-body interactions such as e-e, e-ph, and ph-e not only characterize the materials but may define symbiotic behaviors paramount to the performance of the devices. Therefore, predicting electrical, thermal, and phonon-driven phenomena at the nanoscale is one of the essential avenues for future applications of the MC approach developed herein that may shed light on the functionality of different devices, otherwise challenging to scrutinize experimentally.

# 6 Statements and Declarations

Competing Interests

The authors declare that they have no conflict of interest.

**Author contributions**

All authors have contributed equally to the work reported in this paper.

**Acknowledgements**

This work was supported by the Center for Thermal Energy Transport under Irradiation and the Energy Frontier Research Center funded by the U.S. Department of Energy, Office of Science, Office of Basic Energy Sciences.

**Data availability statements**

The datasets generated during and/or analyzed during the current study are available from the corresponding author on reasonable request. This manuscript has no associated data or the data will not be deposited.